\documentclass[useAMS,usenatbib,psfig]{mn2e}

\usepackage{graphicx}

\title[SPH simulations of Shakura-Sunyaev instability at intermediate accretion rates]
{SPH simulations of Shakura-Sunyaev instability at intermediate accretion rates}
\author[V. Teresi, D. Molteni, E. Toscano]{V. Teresi$^1$ \thanks{E-mail:
vteresi@unipa.it (VT)}, D. Molteni$^1$ and E. Toscano
$^{1}$ \\
$^{1}$Dipartimento di Fisica e Tecnologie Relative, Universit$\grave{a}$ di Palermo, Viale delle Scienze, Palermo, 90128, Italy\\
}
\begin{document}



\maketitle

\label{firstpage}

\begin{abstract}
We show that a standard Shakura-Sunyaev accretion disc around a
black hole with an accretion rate $\dot{M}$ lower than the
critical Eddington limit does show the instability in the
radiation pressure dominated zone. We obtain this result
performing time-dependent simulations of accretion disks for a set
of values of $\alpha$ and $\dot{M}$. In particular we always find
the occurrence of the collapse of the disc: the instability
develops always towards a collapsed gas pressure dominated disc
and not towards the expansion. This result is valid for all
initial configurations we tested. We find significant convective
heat flux that increases the instability development time, but is
not strong enough to inhibit the disc collapse. A physical
explanation of the lack of the expansion phase is proposed
considering the role of the radial heat advection. Our finding is
relevant since it excludes the formation of the hot comptonizing
corona -often suggested to be present- around the central object
by the mechanism of the Shakura-Sunyaev instability. We also show
that, in the ranges of $\alpha$ and $\dot{M}$ values we simulated,
accretion disks are crossed by significant amplitude acoustic
waves.

\end{abstract}


\section{Introduction}
This work concerns the possibility of occurrence, stated by
Shakura and Sunyaev \citep{b12}, of an instability in the
$\alpha$-disks when the radiation pressure dominates, i.e. in the
so-called A zone. Shakura and Sunyaev demonstrated the existence
of thermal instabilities connected with a difference between
$Q_{\rm -}$ and $Q_{\rm +}$, i.e. the energy emitted per unit area
of the disc and the rate of energy-generation in the disc. The
problem of the existence and outcome of the Shakura-Sunyaev
instability is important in accretion disc physics because it
affects the model of the origin of the comptonization cloud around
some compact objects whose spectra contain a significant part in
the X-ray band. The idea of a hot gas cloud, called
'comptonization cloud', around the disc region close to the
compact object, where photons are pushed towards high frequencies
by the Compton scattering with electrons, is one of the most
common ways to explain the spectral behavior in the X band
\citep{b15}. In general, the outcome of the Shakura-Sunyaev
instability is guessed to be the formation of a hot cloud around
the internal disc region, in which comptonization could happen \citep{b14}.\\
Some authors
already studied the problem of the $\alpha$-disc time evolution.
In 1984 Taam and Lin found that the local Shakura and Sunyaev
stability analysis is confirmed by global time-dependent
simulations of the canonical disc. The instability appears and
gives rise to luminosity fluctuations or bursts \citep{b16}. Taam
and Lin's result is similar to ours despite of the one-dimensional
feature of their simulations (whereas we perform two-dimensional
axisymmetric simulations). In 1987 Eggum et al. found that an
$\alpha$-disc, at sub-Eddington accretion rates, develops the
Shakura-Sunyaev instability and, as a consequence of
that, collapses in a cold thin sheet \citep{b5}.
In 1998 Fujita and Okuda simulated a radiation pressure dominated
$\alpha$-disc in the subcritical accretion rate regime and found
that it is thermally stable \citep{b6}. They analyzed the disc
structure and found that it corresponds to the configuration of
the slim accretion disc model \citep{b1}. In this model, even if
the disc is radiation pressure and electron-scattering dominated,
the equilibrium structure is thermally stable, as a consequence of
an advective radial heat flux that is dominant with respect to the
vertical radiative heat flux. The cases we present are
different, since the structure of the disks we simulated is close
to the $\alpha$-model configuration, rather than to the slim
accretion disc structure.
\\ In all the
mentioned 2D works the evolution of the system has been simulated
only for short time-scales, so their authors cannot exclude they
detect only a transient behavior.\\
In 2001 Agol et al., by performing local two-dimensional
hydrodynamic simulations, showed that, for vertically integrated
dissipation rate proportional to the vertically integrated total
pressure, i.e. in the Shakura-Sunyaev model hypothesis, the
thermal instability develops \citep{b2}. The result of this
instability is generally the disc collapse. Only if a strong
enough initial increase of the local radiation energy density is
given to the disc, there is an expansion, that the authors cannot
follow because a large amount
of matter is pushed out of the simulation region.\\
In our simulations we confirm the results obtained by Eggum and by
Agol. Furthermore our studies can follow the system evolution on a
much larger time-scale and therefore give stronger relevance to
the collapse result. We can also say that the post collapse phase
is very long, at least $4.5 \ 10^{5} \frac{R_g}{c}$ (where $R_g$ is the
Schwartzschild gravitational radius of the black hole and $c$ is
the light speed). We also suggest
that the convective (in $z$ and $r$ directions) energy transport
in the disc may have a twofold role: 1) to increase the time-scale
of the collapse instability; 2) to inhibit the expansion
instability.

\section[]{The physical model}
The time dependent equations describing the physics of accretion
disks are well known. They include: \\
mass conservation

\begin{equation}
\frac{D\rho}{Dt} = -\rho \hspace{0.20cm}
div\stackrel{\rightarrow}{v}
\end{equation}

radial momentum conservation

\begin{equation}
\rho \frac{Dv_r}{Dt} = -\rho \frac{\lambda^2}{r^3} + g_r + (div
\stackrel{\leftrightarrow}{\sigma})_r + f_r
\end{equation}

vertical momentum equation

\begin{equation}
\frac{Dv_z}{Dt} = -\frac{1}{\rho} \frac{dP}{dz} - g_z + f_z
\end{equation}

energy balance

\begin{equation}
\rho\frac{D}{Dt}(E+\frac{1}{2}v^2) =
 \stackrel{\rightarrow}{v} \cdot \stackrel{\rightarrow}{F}\rho +
div (\stackrel{\rightarrow}{v}\stackrel{\leftrightarrow}{\sigma})
-div \stackrel{\rightarrow}{F}
\end{equation}

Here $\frac{D}{Dt}$ is the comoving derivative,
$\stackrel{\rightarrow}{F}$ is the radiation flux, given by:

\begin{equation}
\stackrel{\rightarrow}{F} = - \frac{c}{3\rho(k + \sigma_T)}
\stackrel{\rightarrow}{\nabla}E_{rad}
\end{equation}

$k$ and $\sigma_T$ are the free-free absorption and Thomson
scattering coefficients, $E_{rad}$ is the radiation energy per
unit volume, $\stackrel{\rightarrow}{f}$ is the radiation force,
given by:

\begin{equation}
\stackrel{\rightarrow}{f} = \rho \frac{k + \sigma_T}{c}
\stackrel{\rightarrow}{F}
\end{equation}

$\lambda$ is the angular momentum per unit mass, $E = \epsilon +
\frac{E_{rad}}{\rho}$ is the total internal energy per unit mass,
including gas and radiation terms,
$\stackrel{\leftrightarrow}{\sigma}$ is the viscosity stress
tensor. The component of $\stackrel{\leftrightarrow}{\sigma}$ that
is important in accretion disks is the $r$-$\phi$ one, given by:

\begin{equation}
{\sigma}_{r\phi} = \nu \rho r \frac{d\omega}{dr}
\end{equation}

$\nu = \alpha v_s H$ is the kinematic viscosity, $\alpha$ is the
viscosity parameter of the Shakura-Sunyaev model, $v_s$ is the
local sound speed, $H$ is the disc vertical thickness, $\omega$ is
the local angular velocity and the other terms have the usual gas
dynamic meaning.\\ The gravitational force produced by the black
hole is given by the well-known pseudo-newtonian formula by
\citet{b10}:

\begin{equation}
\vec{F} = -\frac{GM}{(R-R_{g})^{2}}\frac{\vec{R}}{R}
\end{equation}

where $R_g$ is the Schwartzschild gravitational radius of the
black hole, given by:

\begin{equation}
R_g = \frac{2GM}{c^2}
\end{equation}

and $M$ is the black hole mass.\\
\\
We adopt the local thermal equilibrium approximation for the
radiation transfer treatment. However this assumption does not
affect our conclusions.

\section{The numerical method and the simulations performed}
We set up a new version of the Smoothed Particles Hydrodynamics
(SPH) code in cylindrical coordinates, for axis symmetric
problems. We remind that SPH is a lagrangean interpolating method.
Recently it has been shown it is equivalent to finite elements
with sparse grid nodes moving along the fluid flow lines
\citep{Dilts}. For a detailed account of the SPH algorithm see
\citet{b9}. For cylindrical coordinates implementation see
\citet{moltbook,moltchakra}. Our code includes viscosity and
radiation treatment. Let us note that -in general- a lagrangean
code is better suited to capture convective motions than eulerian
codes. With the same spatial accuracy (cell size equal to particle
size) the SPH particle motion is tracked with great accuracy, i.e.
the particle size may be large but its trajectory can be still
determined 'exactly'. To integrate the energy equation we adopted
the splitting procedure. In the LTE condition the radiation energy
density changes according to the well known diffusion equation
given by:

\begin{equation}
\frac{\partial E_{rad}}{\partial t}=
-div\stackrel{\rightarrow}{F}= \stackrel{\rightarrow}{\nabla}
\cdot \left( \frac c{3\rho \kappa_{tot}
}\stackrel{\rightarrow}{\nabla} E_{rad}\right)
\end{equation}

where $k_{tot} = k + \sigma_T$.

In cylindrical coordinates $r,z$:

\begin{equation}
\frac{\partial E_{rad}}{\partial t}=\frac cr\frac \partial {\partial r}\left(
\frac r{3\rho \kappa_{tot} }\frac{\partial E_{rad}}{\partial r}\right) +\frac cr\frac
\partial {\partial z}\left( \frac r{3\rho \kappa_{tot} }\frac{\partial E_{rad}}{%
\partial z}\right)
\end{equation}

The SPH-version of the radiation transfer term is given following
the criteria by \citet{brookshaw}. The cylindrical coordinate
version is given by:

\begin{equation}
\left( \frac{\partial E}{\partial t}\right) _i=\frac 1{r_i}\sum_{j=1}^N\frac{%
m_j}{r_j}\left( \frac{E_i-E_j}{\rho _j}\right) D_{ij}\frac{{\bf R}_{ij}}{%
R_{ij}^2}\cdot \stackrel{\rightarrow}{\nabla} _iW_{ij}
\end{equation}

where for clarity we did not put the subscript $rad$ in $E_{rad}$
and where:

\begin{equation}
D_{ij}=\left( \frac{cr_i}{3\rho _i\kappa _{tot_i}}+\frac{cr_j}{3\rho _j\kappa _{tot_j}}%
\right) \ \ ,\ \ \ {\bf R}_{ij}=(r_i-r_j,z_i-z_j)
\end{equation}

This formula can be obtained by the same procedure explained by
Brookshaw, but taking into account that -in cylindrical
coordinates- the particles masses are defined as $m_k=2\pi \ \rho
_k\ r_k\ \Delta r_k\ \Delta z_k$, that explains the further
division by $r_j$ in the term $\frac{m_j}{r_j}$ .

The reference units we use are $R_g$, for length
values, and $R_g/c$ for time values. \\
We performed several simulations, the ones commented here had the
following parameter values:

a)     $\alpha = 0.01$,    $\dot{M} = 0.7$,  domain $R_{1}-R_{2} =
3-300$, $h = 0.3$;

b)     $\alpha = 0.1$,     $\dot{M} = 0.3$,  domain $R_{1}-R_{2} =
3-200$, $h = 0.5$;

c)     $\alpha = 0.01$,    $\dot{M} = 0.7$,  domain $R_{1}-R_{2} =
42-58$, $h = 0.04$;

where $\dot{M}$ is in units of $\dot{M}_E$ and $\dot{M}_E$ is the
critical accretion rate. For all cases the central black hole mass
is $M = 10 \ M_{\odot}$ The spatial resolution we adopt is $h$. In
the case 'a' we have $N = 29704$ particles at time $t = 0$.\\

We used a variable $h$ procedure \citep{hvar}. The $h$ values
above reported are the initial ones. In our procedure, in order to
have a not too small particle size (and therefore not too great
CPU integration times), we put a floor for the $h$ values: $h$ is
chosen as the maximum between the value given by the variable $h$
procedure itself and $1/10$ of the disc vertical half thickness.
So we have nearly 10 particles along the disc half thickness even
in the collapsed region. This floor was not adopted for the case
'c' since we already have a good resolution with a not exceeding
number of particles ($N=67518$).

The boundary conditions of the simulations are not fixed: as the
SPH particles move around, the simulation region follows the form
assumed by the disc and the values of the physical variables at
the boundary of the disc are the values that
characterize the boundary particles at a certain time.\\
The spatial extension of the initial configuration is decided
by establishing a radial range of physical interest and a vertical
extension given by the disc thickness of the Shakura-Sunyaev model.\\
For radiation, the boundary conditions we used are based on the
assumption of the black-body emission and particularly on the
Brookshaw approximation \citep{brookshaw}. At every time step
boundary particles are identified by geometrical criteria (the
particle having the maximum absolute value of $z$ in a vertical
strip of radial width given by h is a boundary particle). The
boundary particle loses its thermal energy according to the
formula given by Brookshaw (that is an approximation of the
diffusion equation at the single particle level), that states the
particle cooling rate proportional to

\begin{equation}
\frac{QT}{h^2}
\end{equation}

where

\begin{equation}
Q = \frac{4acT^3}{3\rho \kappa_{tot}}
\end{equation}

In all our simulations the boundary particles never reach an
optical thickness lower than 10.

Let us now comment the results of our study referring to the
figures of the simulation data.

\begin{figure}
\begin{center}
\includegraphics[scale = 0.35,angle = 270.]{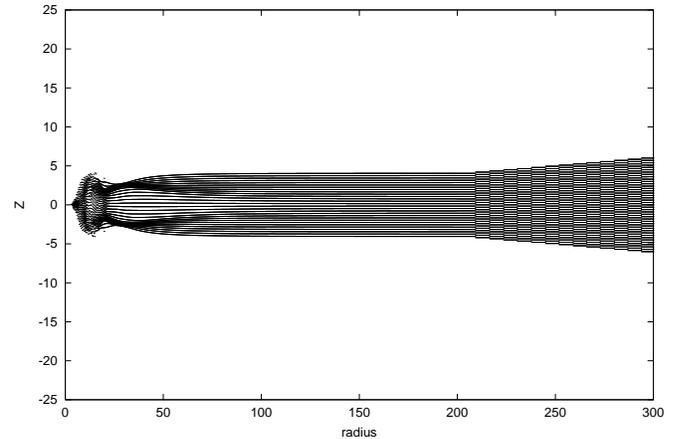}
\caption{The $r$-$z$ profile of the disc of case 'a' is shown at
the time $t = 282 \ R_g/c$ . Every SPH particle is represented by
a small dot. On the $x$ axis the $r$ values in units of $R_g$ are
represented. On the $y$ axis the $z$ values in units of $R_g$ are
represented.}
\end{center}
\end{figure}

Fig. 1 shows the disc structure for the case 'a' at the
adimensional time $t = 282$. The disc profile shows the Z height
of the disc, constant along the A zone. In this figure it is
evident an initial large convective motion in the disc in a region
close to the black hole.

\begin{figure}
\begin{center}
\includegraphics[scale = 0.35,angle = 270.]{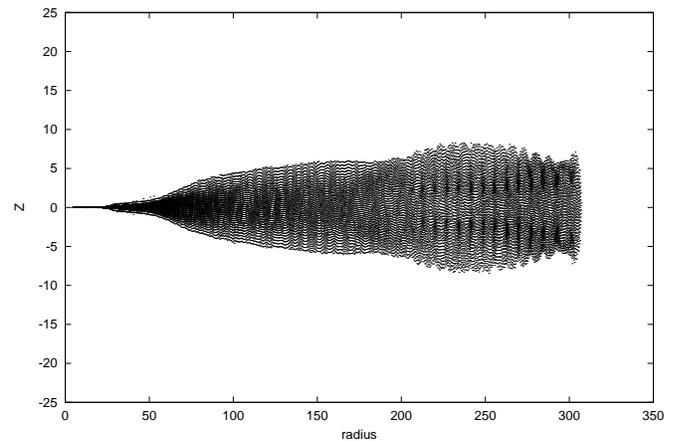}
\caption{The $r$-$z$ profile of the same disc case 'a' at the time
$t = 12000 \ R_g/c$ is shown.}
\end{center}
\end{figure}

Fig. 2 shows the same disc at the larger time $t = 12000$. The
collapsed zone is clearly shown.

\begin{figure}
\begin{center}
\includegraphics[scale = 0.35,angle = 270.]{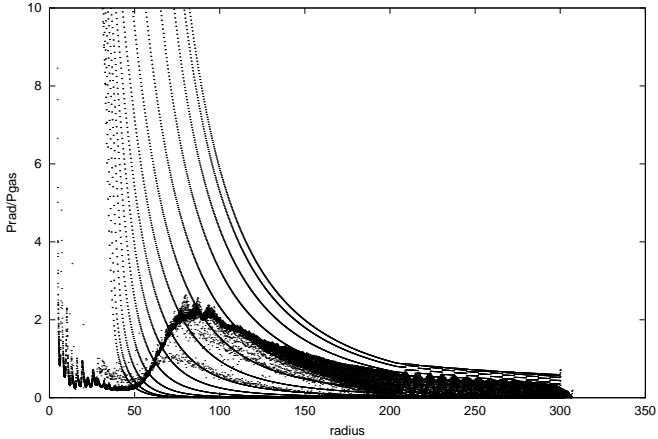}
\caption{The ratio $P_{rad}/P_{gas}$ at the times $t = 282$ and $t
= 12000$ is shown. On the $x$ axis the $r$ values in units of
$R_g$ are represented. The configuration at the later time,
collapsed, exhibits a gas pressure dominated zone up to $r = 70
R_g$, whereas the other configuration, at a earlier time, shows
that, up to $150 \ R_g$, the disc is radiation pressure
dominated.}
\end{center}
\end{figure}

In Fig. 3, the ratio $P_{rad}/P_{gas}$ at the times $t = 282$ and
$t = 12000$ is shown, exhibiting evidence of the collapse.

\begin{figure}
\begin{center}
\includegraphics[scale=0.35,angle=270.]{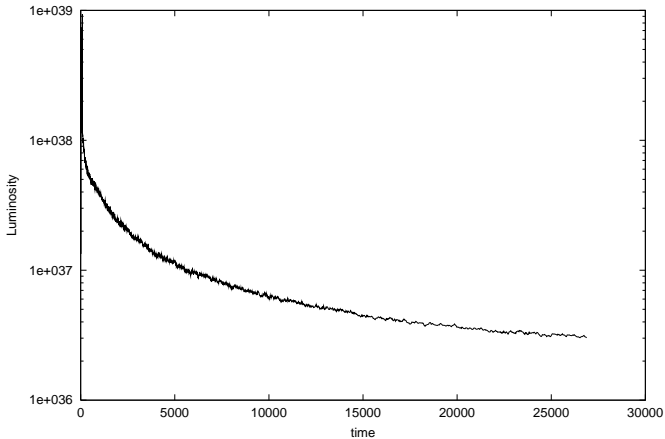}
\caption{The time behavior of the disc luminosity is shown. On the
$x$ axis the time values in units of $R_g/c$ are represented. On
the $y$ axis the luminosity values in units of $erg
\hspace{0.20cm} sec^{-1}$ are represented.}
\end{center}
\end{figure}

Fig. 4 shows that the disc luminosity is steadily decreasing from
the initial theoretical value to much lower values due to the
lower temperature reached by the collapsed A zone.\\

Case 'b' has a larger $h$ and a lower number of particles and it
was possible to follow on the simulation up to the large time $t =
680000$. The $r-z$ distribution of the disc particles is very
similar to case 'a'.

\begin{figure}
\begin{center}
\includegraphics[scale=0.35,angle=270.]{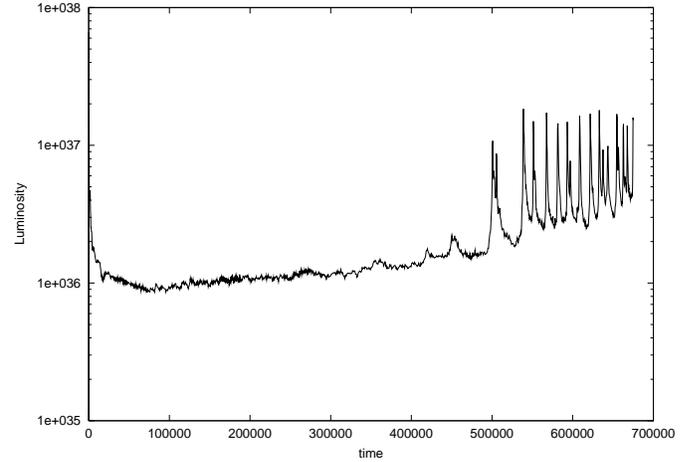}
\caption{The time behavior of the disc luminosity is shown for the
disc of case 'b'.}
\end{center}
\end{figure}

We show in Fig. 5 the luminosity of this disc versus time. It is
apparent that the luminosity has a large decrease during the
collapse going down from a peak at $L = 6.5 \ 10^{37} \ erg
\hspace{0.20cm} sec^{-1}$ to $L = 8.5 \ 10^{35} \ erg
\hspace{0.20cm} sec^{-1}$. After the collapse the disc luminosity
has a slow increase. Nearly after the time $t = 4.8 \ 10^5$
(corresponding to physical time $t = 48 \ sec$, for our
parameters) the luminosity reached an average value of $L = 3 \
10^{36} \ erg \hspace{0.20cm} sec^{-1}$ and starts to show a
strong flaring activity with luminosity reaching values up to $L =
1.6 \ 10^{37} \ erg \hspace{0.20cm} sec^{-1}$ . Obviously the
possibility of a 'recharge' of the inner zone is to be expected
and has been guessed by Eggum and Agol, but -up to now- it was not
given any estimate of the time-scale involved. It has to be noted
that the refilling time-scale is definitely shorter than the
radial viscous drift time-scale $t_{drift} = r^2 /\nu$ as derived
from the canonical disc structure.

In the 'c' case we simulated a small sector of the disc in the
radiation pressure dominated
 zone with a very accurate spatial resolution. This case is similar to the Agol's
case 1 \citep{b2}, with nearly the same disc parameters and
numerical resolution (the finest grid used for their case 1 is a
$256\times256$ one, we have 67518 particles). We also obtained a
vertical collapse instability of the disc and no expansion.

\begin{figure}
\begin{center}
\includegraphics[scale=0.35,angle=270.]{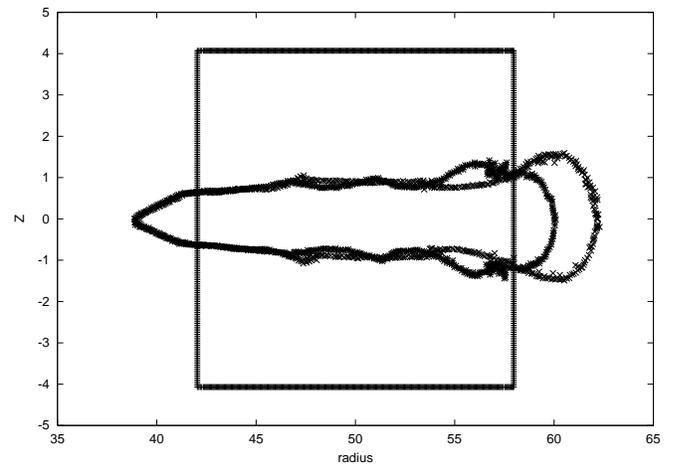}
\caption{The border particles of the disc case 'c' are shown at
the times $t = 0$ (vertical crosses), $t = 20000 \ R_g/c$
(diagonal crosses) and $t = 22400 \ R_g/c$ (asterixes).}
\end{center}
\end{figure}

Fig. 6 shows the  $r-z$ profiles at times $t = 20000$ and $t =
22400$ of the collapsing disc, compared with the disc profile in
the initial configuration. The collapsed structure is completely
gas pressure dominated. In the figure is also apparent the fact
that there are waves travelling along the disc.

During the simulations, we also calculate the $z$-averaged
convective and advective heat fluxes at different values of $r$.
With 'convective flux' we mean the heat flux along the
$z$-direction due to the vertical motion of the gas, whereas with
'advective flux' we mean the heat flux along the $r$-direction due
to the radial motion of the gas. The main time-scales involved in
our problem are defined as follows.\\ The mean convection
time-scale is $t_{conv} = \frac{{\epsilon}H}{F_{conv}}$, where
$F_{conv}$ is the $z$-averaged convective heat flux and $\epsilon$
is the total energy density. The mean radiation diffusion
time-scale is $t_{rad} = \frac{\epsilon_rH}{F_{rad}}$, where
$F_{rad}$ is the $z$-averaged radiative heat flux and $\epsilon_r$
is the radiation energy density. The mean heating time-scale is
$t_{heat} = \frac{dE_{rot}}{\tau_{r\phi}\frac{d\omega}{dr}dr}$,
where $dE_{rot}$ is the rotational energy of the disc ring between
$r$ and $r + dr$, $\tau_{r\phi}$ is the $z$-averaged viscosity
stress and $\omega$ is the local angular velocity of the disc.\\

\begin{figure}
\begin{center}
\includegraphics[scale=0.35,angle=270.]{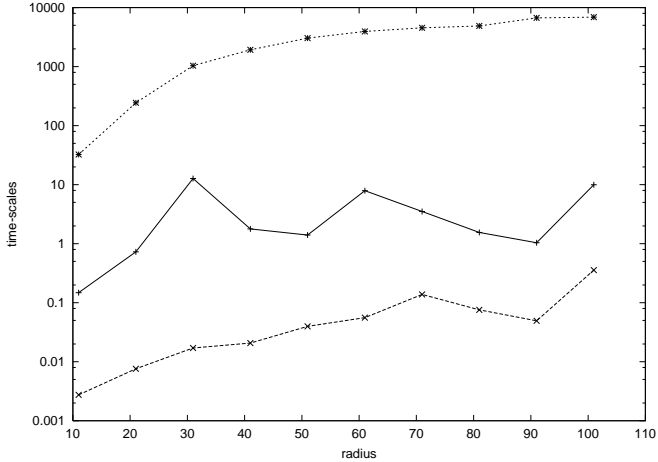}
\caption{The radiative, convective and heating time-scales,
ordered from the bottom to the top of the figure, are shown versus
$r$. On the $y$ axis the time values in units of $R_g/c$ are
represented.}
\end{center}
\end{figure}

Fig. 7 shows the mentioned time-scales evaluated at different
radii $r = 10 - 100 \ R_g$ for case 'b', after time $t = 10^5 $.

From the comparison among the evaluated three time-scales it is
clear that the convection time-scale is  -in general- intermediate
between the other two time-scales, i.e. the convection time-scale
is greater than the vertical radiation diffusion time-scale and
smaller than the heating time. So we can say that the vertical
convective heat transfer may have a significant role in the
vertical redistribution of the generated heat.

\begin{figure}
\begin{center}
\includegraphics[scale=0.35,angle=270.]{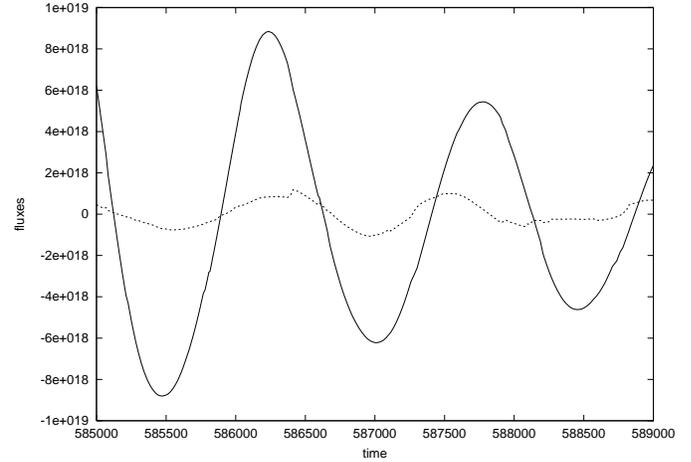}
\caption{The convective and advective fluxes at $r = 30 \ R_g$ are
shown versus time. On the $y$ axis the fluxes values in units of
$erg \hspace{0.20cm} sec^{-1} \hspace{0.20cm} cm^{-2}$ are
represented. The dashed line represents the convective flux
behavior, whereas the continuous one refers to the advective
flux.}
\end{center}
\end{figure}

\begin{figure}
\begin{center}
\includegraphics[scale=0.35,angle=270.]{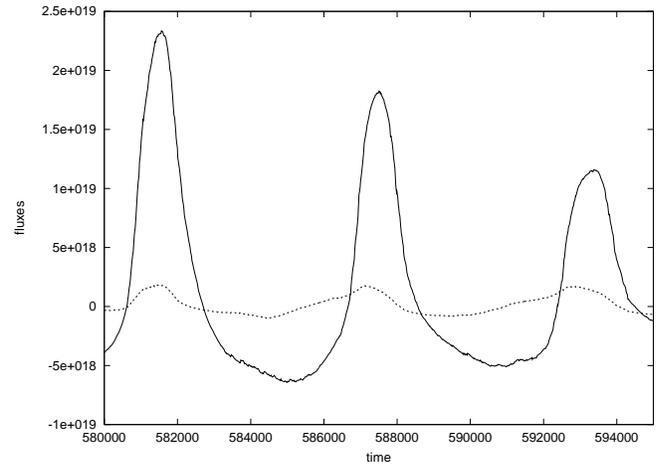}
\caption{The convective and advective fluxes at $r = 80 \ R_g$ are
shown versus time. On the $y$ axis the fluxes values in units of
$erg \hspace{0.20cm} sec^{-1} \hspace{0.20cm} cm^{-2}$ are
represented. The dashed line represents the convective flux
behavior, whereas the continuous one refers to the advective
flux.}
\end{center}
\end{figure}

The role of the advection (radial transport) appears more relevant
than the (vertical) convection. The time variation of the
convective, advective and radiative fluxes is oscillatory and the
oscillation frequency is close to the local keplerian frequency
(Figs. 8 and 9).

The approximate equality between the flux oscillation and
keplerian rotation frequencies is explainable if one assumes that
the flux oscillation is due to the propagation of an acoustic wave
in the disc. Milsom and Taam already showed that accretion disks
are crossed by acoustic waves \citep{b7,b8}. In the limit of
negligible viscosity and advection and for propagation direction
perpendicular to the disc rotation axis, it is possible to have an
analytical formula for the wave frequency. The dispersion relation
for these waves is given by $\omega^2 = k^2 + K^2 v_s^2$
\citep{b4}, where $\omega$ is the angular frequency of the wave,
$K$ is the wave number and $k$ is the so-called local epicycle
frequency, given by $k^2 = 2 \frac{\Omega}{r} \frac{d}{dr}(r^2
\Omega)$, where $\Omega$ is the local angular velocity of the
disc. $k$ is in general close to the local keplerian angular
velocity. Therefore the fact that the waves frequencies and the
keplerian angular velocities are close is a suggestion of the
acoustic origin of such waves in accretion disks.

Finally, the waves we obtain have amplitudes that remain roughly
constant in time. No other kind of disc structure variability is
present.

We may add that the wave phenomenon appears -at different
intensity levels- in all cases we examined. We never obtained a
monotonic regular radial speed as the Shakura-Sunyaev model
predicts.

\section{Discussion}
Here we discuss two items : the absence of the expansion
instability and the refilling of the inner zone of the disc.

The origin of the instability in radiation pressure dominated zone
is clear, but it is not clear why the preferred evolution is
towards the collapsed state and not towards the expanded one. At
subcritical accretion rates, i.e. in our accretion rate regimes,
the expansion instability never occurs. We suggest that this
result can be due to two effects: the enhanced cooling in the
expansion branch and the significant role of advection.

If the disc evolved towards the expansion instability the disc
density would go down and therefore also its optical thickness.
The basic LTE approximation, under which the disc model is built,
breaks down; the disc can then radiate its energy content more
quickly than it is heated by its viscosity. In our view the role
of convection and advection may also contribute to reduce the
expansion instability, but not the collapse instability. This
mechanism has been considered capable to reduce the
Shakura-Sunyaev instability. It is known that such a transfer
should be present in accretion disks \citep{b3,b6,b13}, as a
result of the entropy gradient associated with the radiation field
in thermal equilibrium with the gas in the accretion disc. This
gradient assumes a very large value when the temperature
distribution is determined only by the radiative heat transfer and
the viscous energy production rate. Following Bisnovatyi-Kogan and
Blinnikov analysis, the vertical convective motion and the
consequent vertical heat transfer have, therefore, the role of
producing an isentropic $z$ structure. If we indicate with
$F_{conv}$ the vertical convective heat flux, the equation of the
vertical heat transfer is, in presence of convection:

\begin{equation}
-\frac{d}{dz} (F_{rad} + F_{conv}) + \eta (r \frac{d\omega}{dr})^2
= 0
\end{equation}

where $F_{rad}$ is the standard radiative flux. So we have a heat
flux whose value has been enhanced from $F_{rad}$ to:

\begin{equation}
F_{tot} = F_{rad} + F_{conv} = F_{rad} (1 +
\frac{F_{conv}}{F_{rad}}) = f F_{rad}
\end{equation}

where

\begin{equation}
f = 1 + \frac{F_{conv}}{F_{rad}}
\end{equation}

It has been demonstrated that this enhancement of the vertical
heat flux by a factor $f$ produces an increase of the instability
development time by the same factor $f$ \citep{b13}. From our
simulations we obtain that, in the parameter range we explored,
$f$ is always less than $2$. Such a small value of $f$ does not
increase significantly the instability development time. However
the advected radial heat is not taken into account in the
calculations of \citet{b3,b13}.

\begin{figure}
\begin{center}
\includegraphics[scale=0.35,angle=270.]{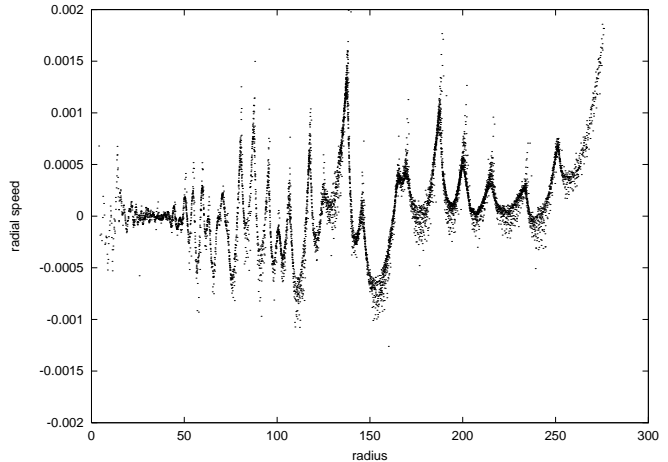}
\caption{The radial speed of the disc parcels is shown versus $r$.
On the $y$ axis the speed values in adimensional units are
represented.}
\end{center}
\end{figure}

\begin{figure}
\begin{center}
\includegraphics[scale=0.35,angle=270.]{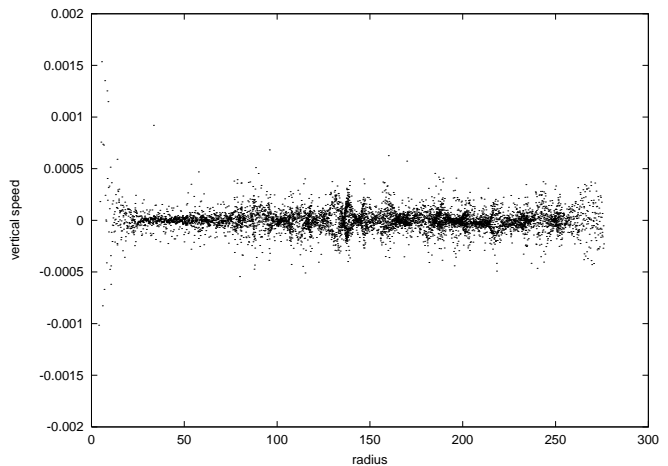}
\caption{The vertical speed of the disc parcels is shown versus
$r$. On the $y$ axis the speed values in adimensional units are
represented.}
\end{center}
\end{figure}

Figs. 10 and 11 show the radial and vertical speeds of the disc
parcels versus $r$. It is clearly apparent from Figs. 8, 9, 10 and
11 that the radial flux is greater than the vertical one.

As the radial heat flux assumes a significant value, the heat
produced at a certain radius is carried away from that point of
the disc where  it had the possibility to produce the expansion
instability. Furthermore this advection helps the cooling of the
disc towards the collapsed configuration of the instability.

We obtain values of the radial heat flux that are large with
respect to the vertical convective flux. Therefore we propose that
the absence of the disc expansion may be due also to the role of
the radial advective heat transfer.\\

 As regards the collapse
instability development times, we have compared the theoretical
values with the ones calculated from the luminosity time behavior:
we see that, when the disc collapses, its luminosity decreases
very quickly and the time on which the luminosity reduction occurs
can be assumed as the instability development time. As an example,
we can see the comparison between the two times in the case 'b'.
The theoretical instability development time, $t_{th}$, can be
calculated from the instability development rate $\Omega_{th}$,
given by $\Omega_{th} = \alpha \omega \frac{6(5\beta_r -
3)}{A(\beta_r)}$, with $A(\beta_r) = 8 + 51\beta_r - 3\beta_r^2$
and $\omega$ equal to the angular velocity of the disc at the
radius considered \citep{b12}. With this expression for
$\Omega_{th}$ we can calculate the instability development time
$t_{th}$ as $t_{th} = \frac{1}{\Omega_{th}}$.
The value of $t_{th}$ at the innermost radii is of the order of $10^{-1} \ sec$.\\
From the luminosity time behavior we deduce similar values. \\
\\

\begin{figure}
\begin{center}
\includegraphics[scale=0.35,angle=270.]{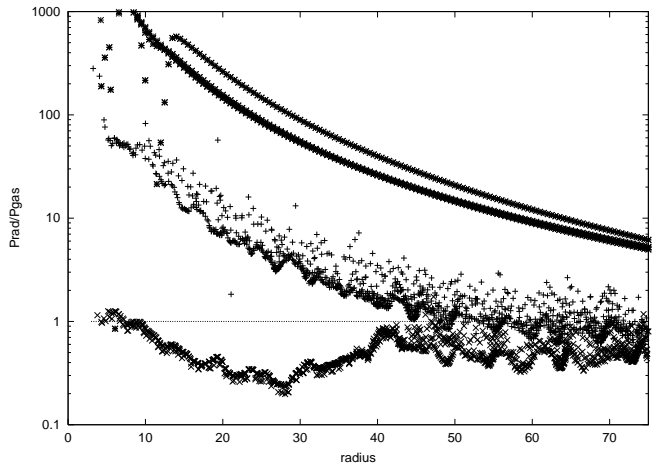}
\caption{The ratio $P_{rad}/P_{gas}$ in the initial configuration
(asterixes) and at the times $t = 10^5$ (diagonal crosses) and $t
= 6 \ 10^5$ (vertical crosses) is shown. The initial
configuration, in the shown radial range, is radiation pressure
dominated. At the intermediate time ($t = 10^5$) the disc,
collapsed, is gas pressure dominated. The configuration at the
later time ($t = 6 \ 10^5$), refilled, shows again a radiation
pressure dominated zone up to $r = 60 \ R_g$.}
\end{center}
\end{figure}

Fig. 12 shows the ratio of the radiation pressure to the gas
pressure versus the radial distance for the case 'b' when the disc
was in the initial configuration, in the collapsed state (time $t
= 1\ 10^5$) and when it was in the refilled state (time $t = 6\
10^5$). It is clear that, in the A zone, the refilled state has a
$P_{rad}/P_{gas}$ value larger than 1, but smaller than the very
initial value, at time $t = 0$. The presence of this zone with
$P_{rad}
> P_{gas}$ also in the refilled configuration suggests that the
Shakura-Sunyaev instability may be again operating to produce the
recurrent flaring activity.

The refilling time-scale computed from the simulation data is
about $5\ 10^{5}$, that is shorter than the theoretical drift
time-scale due to the viscosity. We argue that the refilling is
driven not only by the viscosity: also the wave phenomenon may
play significant role. Indeed  the plain theoretical model
predicts a time-scale  $t_{drift} = r^2 /\nu = 1.8 \ 10^7$. This
fact should be taken into account developing theoretical models of
recurrent flaring disc activity due to refilling.

\section{Conclusions}

The results of our simulations of the time evolution of $\alpha
P_{tot}$ disks, with a large portion in radiation pressure
dominating conditions, show definitely that the Shakura-Sunyaev
collapse instability develops. After a long time the disc recovers
partially its luminosity and furthermore it exhibits a flaring
like activity. The recovery time-scale is shorter than the viscous
drift time-scale. We attribute to the advective - convective heat
transfer a significant role to determine both the recovery and the
not occurence of the expansion instability.

We suggest that the reason for which we see no expansion
instability is the presence of a radial convective motion, that
isn't considered in the Shakura and Sunyaev analysis, but is
naturally simulated by our code. In such conditions, the radial
heat transfer due to the advection of matter (and thermal energy)
carries the excess of energy produced by viscosity (that would
cause the thermal expansion) away from the disc element at that
$r$. These results imply some problem for the model of the
formation of the comptonization cloud supposed to exist in many
disc configurations to explain the observed radiation spectra.
From the canonical disc it seems impossible to produce that
comptonization cloud via the plain Shakura-Sunyaev instability.

Another relevant aspect of our simulations is the presence of
acoustic waves; they can cause a kind of periodical variability of
the radiation spectrum and intensity that could be related, from
the observational point of view, with the phenomenon of QPOs
\citep{psaltis}. Oscillation frequencies of the radiative flux we
calculated are compatible with some QPOs. It is true that the
total disc luminosity takes into account the contributions from
the whole disc and so the wave phenomenon may not appear. However
one should take also into account the possibility that the
reflection of the radiation coming from inner disc zones by outer
zone waves could produce an enhanced oscillation in the total
luminosity. Such a study requires an accurate treatment of the
interaction between the emitted radiation and the disc itself and
of the radiation transfer in the outermost layers of the disc,
that is beyond this contribution.

The flaring activity could be also responsible of QPO phenomena.
We are planning to confirm and study its characteristics by
numerical simulations with increased spatial resolution.

%

\newpage
{\bf{Figure captions}}

Fig. 1. The $r$-$z$ profile of the disc of case 'a' is shown at
the time $t = 282 \ R_g/c$ . Every SPH particle is represented by
a small dot. On the $x$ axis the $r$ values in units of $R_g$ are
represented.
On the $y$ axis the $z$ values in units of $R_g$ are represented.\\
Fig. 2. The $r$-$z$ profile of the same disc case 'a' at the time
$t = 12000 \ R_g/c$
is shown.\\
Fig. 3. The ratio $P_{rad}/P_{gas}$ at the times $t = 282$ and $t
= 12000$ is shown. On the $x$ axis the $r$ values in units of
$R_g$ are represented. The configuration at the later time,
collapsed, exhibits a gas pressure dominated zone up to $r = 70 \
R_g$, whereas
the other configuration, at a earlier time, shows that, up to $150 \ R_g$, the disc is radiation pressure dominated.\\
Fig. 4. The time behavior of the disc luminosity is shown. On the
$x$ axis the time values in units of $R_g/c$ are represented.
On the $y$ axis the luminosity values in units of $erg \hspace{0.20cm} sec^{-1}$ are represented.\\
Fig. 5. The time behavior of the disc luminosity is shown for the disc of case 'b'.\\
Fig. 6. The border particles of the disc case 'c' are shown at the
times $t = 0$ (vertical crosses), $t = 20000 \ R_g/c$ (diagonal
crosses)
and $t = 22400 \ R_g/c$ (asterixes).\\
Fig. 7. The radiative, convective and heating time-scales, ordered
from the bottom to the top of the figure, are shown versus $r$.
On the $y$ axis the time values in units of $R_g/c$ are represented.\\
Fig. 8. The convective and advective fluxes at $r = 30 \ R_g$ are
shown versus time. On the $y$ axis the fluxes values in units of
$erg \hspace{0.20cm} sec^{-1} \hspace{0.20cm} cm^{-2}$ are
represented.
The dashed line represents the convective flux behavior, whereas the continuous one refers to the advective flux.\\
Fig. 9. The convective and advective fluxes at $r = 80 \ R_g$ are
shown versus time. On the $y$ axis the fluxes values in units of
$erg \hspace{0.20cm} sec^{-1} \hspace{0.20cm} cm^{-2}$ are
represented.
The dashed line represents the convective flux behavior, whereas the continuous one refers to the advective flux.\\
Fig. 10. The radial speed of the disc parcels is shown versus $r$.
On the $y$ axis the speed values in adimensional units are represented.\\
Fig. 11. The vertical speed of the disc parcels is shown versus
$r$.
On the $y$ axis the speed values in adimensional units are represented.\\
Fig. 12. The ratio $P_{rad}/P_{gas}$ in the initial configuration
(asterixes) and at the times $t = 10^5$ (diagonal crosses) and $t
= 6 \ 10^5$ (vertical crosses) is shown. The initial
configuration, in the shown radial range, is radiation pressure
dominated. At the intermediate time ($t = 10^5$) the disc,
collapsed, is gas pressure dominated. The configuration at the
later time ($t = 6 \ 10^5$), refilled,
shows again a radiation pressure dominated zone up to $r = 60 \ R_g$.\\

\bsp

\label{lastpage}

\end{document}